\documentclass[12pt,preprint]{aastex}
\usepackage{graphicx}
\usepackage{times}
\usepackage{float}
\usepackage{wrapfig}
\usepackage{float}
\usepackage{rotating}
\usepackage{lscape}
\usepackage{longtable}

\shorttitle{Debris Disks at 670 Myr} 
\shortauthors{Urban \etal (2011)}
                                                                                
\begin{document}
\title{The Incidence of Debris Disks at 24 $\mu$m and 670 Myr}
\author{Laurie E. Urban\altaffilmark{1,2}, George Rieke\altaffilmark{3}, Kate Su\altaffilmark{3}, David E. Trilling\altaffilmark{1}}
\altaffiltext{1}{Department of Physics and Astronomy, Northern Arizona
  University, P.O. Box 6010, Flagstaff, AZ 86011, USA}
\altaffiltext{2}{Institute for Astronomy, University of Hawaii at
  Manoa, 2680 Woodlawn Dr, Honolulu, HI 96822, USA}
\altaffiltext{3}{Steward Observatory, University of Arizona, Tucson, AZ 85721, USA}

\begin{abstract}

We use {\em Spitzer Space Telescope} 24~$\micron$ data to search for
debris disks among 122~AFGKM stars from the $\sim$ 670~Myr clusters Hyades,
Coma Ber, and Praesepe, utilizing a number of advances in data
reduction and determining the intrinsic colors of main sequence stars.
For our sample, the 1$\sigma$ dispersion about the main sequence
V-K$_S$, K$_S$-[24] locus is approximately 3.1\%.  We identify
seven debris disks at 10\% or more ($\ge$ 3$\sigma$ confidence level)
above the expected K$_S$ - [24] for purely photospheric emission. The incidence of excesses of 10\% or 
greater in our sample at this age is $5.7^{+3.1}_{-1.7}$\%. Combining with results from the literature, 
the rate is $7.8^{+4.2}_{-2.1}$\% for early-type (B9 - F4) stars and $2.7^{+3.3}_{-1.7}$\%  for solar-like (F5 - K9) stars. 
Our primary sample has strict criteria for inclusion to allow
comparison with other work; when we relax these criteria, three
additional debris disks are detected. They are all around stars of
solar-like type and hence reinforce our conclusion that disks around
such stars are still relatively common at 670 Myr and are similar to the rate around early-type stars. The apparently small difference in
decay rates between early-type and solar-like stars is inconsistent with the first order theoretical predictions that
the later type stellar disks would decay an order of magnitude more quickly than the earlier type ones. 

\end{abstract}

\section{Introduction}
Most stars form with an accompanying protoplanetary disk.  Initially
these disks begin as dense optically thick regions mostly made up of
primordial gas and dust; these constituents make the disks readily detectable 
in thermal continuum and multiple emission lines. These disks provide the environment 
for massive planet formation. Ironically, once planets have formed 
the disks disappear after a few million years (e.g.,
Haisch et al.~2001).  The resulting planetary systems remain difficult to detect directly.
The systems have not ceased evolving at this point; terrestrial planets 
continue to grow from the solid remnants of the disks (e.g., Agnor,
Canup \& Levison 1990; Chambers 2001; Raymond et al. 2006;
Morishima et al. 2010), and massive planets may migrate
inward or outward (Hahn \& Malhotra 1999; Gomes, Morbidelli \& Levison
2004; Levison et al. 2007). These processes are marked by dynamical stirring of the asteroidal
or planetesimal bodies, which are pulverized into dust that is warmed by the star and glows in the infrared.
This dust dissipates rapidly due to radiation pressure,
Poynting-Robertson drag, or collisional destruction (e.g., Lagrange et
al. 2000; Dominik \& Decin 2003).  Hence, these debris disks must reflect the current
state of the systems, and the infrared excesses 
provide snapshots of the state of the circumstellar material and of the
degree of dynamical activity causing it to undergo collisions. 
A series of these snapshots gives a view of evolution of the planetary system 
(e.g., Kenyon \& Bromley 2004; Wyatt
2008). 

Debris disks were first observed in the infrared with the {\em Infrared
Astronomical Satellite (IRAS)} (Aumann et al.\ 1984; see also Rhee et
al.\ 2007). Many subsequent debris disk observations were made with
the {\em Infrared Space Observatory (ISO)} (e.g., Decin et al.\ 2000;
Spangler et al.\ 2001; Habing et al.\ 2001; Decin et al.\ 2003).
The Multiband Imaging Photometer for {\em Spitzer} (MIPS: Rieke et al.\ 2004)
on board the {\em Spitzer Space Telescope} provided another significant
increase in sensitivity, observing a large number of debris disks
at 24~and 70~$\micron$. {\it Herschel} is currently extending our understanding of large samples 
of debris disks to longer infrared and submillimeter wavelengths.

The evolution of the 24 \micron ~emission of debris disks is
particularly interesting because this wavelength traces inner
zones of these systems that may reveal processes near the ice line
(Morales et al.~2011) and because the dynamical timescale for these zones is shorter
than for those probed at longer wavelengths, so the evolution should proceed
relatively rapidly. Previous surveys have verified that the 24 $\mu$m emission of debris disks
decays significantly over time. About 50\% of early-type stars younger than 30
Myr have debris disks detected at 24 $\micron$ (Rieke et al.\ 2005 and
Su et al.\ 2006). However, at 100 to 115 Myr the detection rate has
dropped to about 35\% (see below).
G\'asp\'ar et al. (2009) find the Praesepe cluster at $\sim$ 750 Myr to
have an excess rate of only $\sim$ 2\%.  Surveys of stars older than $\sim$ 1
Gyr report an excess rate of $\sim$4\% or less (Meyer et al. 2008,
Trilling et al. 2008, Koerner et al. 2010).  

A better understanding of the decay process will help build a picture
of the evolution of planetary systems.  Toward this goal, we examine
MIPS data at 24 $\mu$m for three stellar clusters with ages
$\sim$670~Myr: Hyades ($\sim$ 650 Myr; Perryman et al. 1998, De Gennaro et al. 2009), 
Coma Ber ($\sim$ 600 Myr; Collier et al. 2009), and Praesepe ($\sim$ 750 Myr; G\'asp\'ar et al. 2009). We improve on previous
studies by combining results for three
separate clusters of similar ages. This study also improves over similar work on field stars
because the ages of our sample are much better constrained. Combining
three clusters provides a high fidelity measurement over a large range of spectral types
from early K for the closest cluster (Hyades) to A-types
with reasonable statistics for the richest but most distant
(Praesepe). 

We discuss in Section 2 how we selected our sample for this study,
reduced the data, excluded stars with large errors or bad
measurements, and placed our retained targets on a $V-K_S$ versus $K_S-[24]$ color-color diagram (Gorlova et al.\ 2006). We
improve on previous results by utilizing a new locus for the photospheres of main sequence stars. In
Section 3 we determine which stars have excess emission at
24~$\micron$ and compare our work with that of others. Section 4
discusses the excess rate at 670~Myr as a function of spectral
type. We also compare these results to previously published excess
rates and discuss the decay rate of 24 $\micron$ excesses.  Our conclusions are in Section 5.

\section{Observations and Data Reduction}

\subsection{Sample Selection and Data Reduction}

We searched the {\it Spitzer} archive for observations of stars in published membership lists for the Hyades, Coma Ber,
and Praesepe. Our initial samples included 78~stars
from the Hyades (Paulson et al.\ 2004), 84~from Coma Ber (Abad et
al.\ 1999), and 193~from Praesepe (G\'asp\'ar et al.\ 2009). 
Some of the stars have accompanying 70 $\mu$m
observations; here we only report the 24 $\mu$m results, 
some of which have been published previously
(Rieke et al. 2005; Su et al. 2006; Cieza et al. 2008; 
Carpenter et al. 2008; G\'asp\'ar et al. 2009). 

For consistency, we
re-processed all the data as part of a {\it Spitzer} legacy catalog
(Su et al. 2010), using the MIPS instrument team Data Analysis Tool (Gordon et
al.~2005) for basic reduction. In addition, a second flat field
constructed from the 24 $\mu$m data itself was 
applied to all the 24 $\mu$m results to remove scattered-light gradients
and dark latency (e.g., Engelbracht et al. 2007).
The processed data were then combined using the World Coordinate System (WCS)
information to produce final mosaics with pixels half the size of the
physical pixel scale. The majority of the stars in the Praesepe
cluster were observed in scan-map mode as presented in
G\'asp\'ar et al. (2009). We used the same data reduction as in that paper, but
did the photometry differently as described below.

We extracted the photometry using PSF fitting. The input PSFs were
constructed using observed calibration stars and smoothed STinyTim
model PSFs, and have been tested to ensure the photometry results are
consistent with the MIPS calibration (Engelbracht et
al.\ 2007). Aperture photometry was also performed, but the results 
were only used as a reference to screen targets that might have
contamination from nearby sources or background nebulosity.

The random photometry errors were estimated based on the
pixel-to-pixel variation within a 2\arcmin$\times$2\arcmin~box centered
on the source position. The final photometry errors also included the
errors from the detector repeatability ($\lesssim$1\% at 24 $\mu$m,
Engelbracht et al.~2007). The measured flux densities, 24 $\mu$m
magnitudes ([24], using 7.17 Jy as the zero magnitude flux; Rieke et
al.~2008), and associated errors are listed in Table
\ref{Cluster Stars Considered}. 

\subsection{Culling the Sample}

We further trimmed the sample based on the criteria described below to
ensure our analysis only includes measurements of uniform quality. 
Because of {\it Spitzer's} exceptional pointing accuracy
($<$1\arcsec), we considered a source only where the PSF fitting position falls within 1\farcs5 ~of the member position (from the membership catalogs). 
Sources that failed this positional test are noted as ``6'' in Table
\ref{Cluster Stars Considered}. We then eliminated by visual inspection of the 24 $\mu$m
images all stars with a nearby source (within 6\arcsec) of the
target star (noted as ``1'' in Table \ref{Cluster Stars Considered}).  We 
excluded all stars that are classified as giants
(luminosity class III) (noted as ``7'' in Table \ref{Cluster Stars Considered}). We
also excluded all stars that could not be detected by eye, implying a
signal to noise ratio $<$3 (noted as ``3'' in Table
\ref{Cluster Stars Considered}); this step guarded against false detections in regions 
where the background had residual structure.  Finally, we excluded all 
measurements with a FWHM either much greater or less than the nominal FWHM of 
the MIPS beam of $\sim$ 5\farcs5 (noted as ``2'' in Table \ref{Cluster Stars Considered}), 
which is an indication
of confusion with a second object.

In our analysis we rely on $K_S$ magnitudes from 2MASS (Cutri et al.\
2003). For stars in the Hipparcos catalog (Perryman et al.\ 1997), we
adopted the listed Johnson V photometry. For stars that are not in
Hipparcos but are in Praesepe we adopted the V magnitudes gathered
by G\'asp\'ar et al.\ (2009); we collected additional V
magnitudes from the Simbad database. A total of 8 stars have no
available V magnitudes, noted as ``5'' in Table
\ref{Cluster Stars Considered}; these stars are not included in our analysis.

Finally, we removed stars with large errors in $K_S-[24]$. (The errors
from $V-K_S$ are not used to remove stars from our sample since the
errors in the V magnitudes are generally small and any stars with
large errors in K$_S$ will be eliminated because of the errors in K$_S-[24]$.) We took the
$K_S$ errors ($e_{K_S}$) from 2MASS. For the MIPS photometry, we
combined two types of error. The first, {\em e$_{24,pp}$}, is based on
the pixel-to-pixel variation near the source and represents the random
photometric error. The second arises from the
overall repeatability of the MIPS 24 $\mu$m measurement at $\sim$ 1\% of
the source fluxes (Engelbracht et al.\ 2007). The final error
($e_{24}$) in the 24~$\micron$ flux ($f_{24}$) is thus 
\begin{equation}
e_{24}=\sqrt{e_{24,pp}^2+(0.01*f_{24})^2}. 
\end{equation}
We combine the 24~$\micron$ errors with the $K_S$ errors to find the
total (RMS) error on $K_S-[24]$. Only stars with $e_{K_S-[24]}<0.05$
were retained (exclusion note ``4'' in Table
\ref{Cluster Stars Considered}). Our final sample contains 122~stars: 61~from
Hyades, 25~from Coma Ber, and 36~from Praesepe.  We present all 355
stars in Table~\ref{Cluster Stars Considered}, where we indicate the 122 stars
included and the reason the remaining were excluded from analysis.

\subsection{Main Sequence Definition and Scatter}

\begin{figure}
\figurenum{1}
\label{fig1}
\plotone{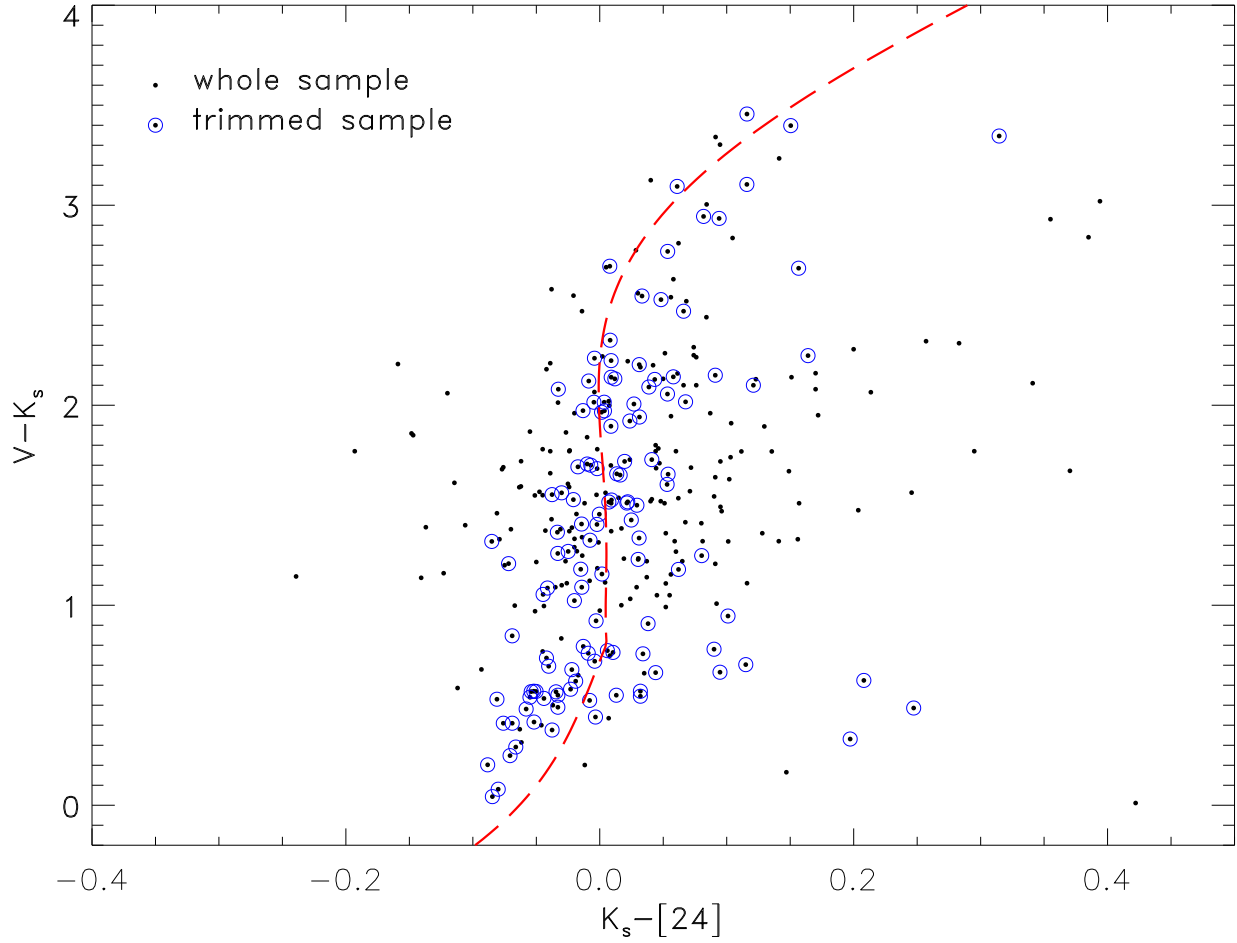}
\caption{$V-K_S$ vs. $K_S - [24]$ color-color plot for 670 Myr cluster members. The entire sample is shown as 
small dots with circles indicating the ones retained in our analysis. The main-sequence locus presented in Eqn (2) and (3) is shown as the dashed line.} 
\end{figure}

Figure \ref{fig1} shows the $V-K_S$ vs. $K_S - [24]$ color-color plot for the whole sample, 
and the final sample retained in our analysis. 
Gorlova et al.\ (2006) introduced such a diagram to determine the photospheric locus for main sequence stars in the Pleaides cluster. They identified stars with infrared excesses as those with $K_S-[24]$ positive by more than 3$\sigma$ relative to the photospheric locus. Here we apply the same general technique. We have used a sample of
$\sim$ 1300 stars from the {\it Spitzer} archive (Su et al. \ 2010) to
derive the locus of main sequence stars in $V-K_S$, $K_S-[24]$
space. 
We define $x=V-K_S-0.8$.  For x$\le$0, we find the following empirical
fits:
\begin{equation}
K_S-[24]=\frac{(50+661x+282x^2+653x^3)}{10000};
\end{equation}
 whereas for $x>0$, we find:
\begin{equation}
K_S-[24] =
\frac{(55-134x+655.5x^2-1095x^3+664.15x^4-145.8x^5+11.01x^6-0.03x^7}{10000}
. 
\end{equation}
At the juncture where $V-K_S=$0 the equations agree so either
of them applies.  The RMS scatter around these fits (determined by
fitting gaussians to the distribution so the result is not biased by
stars with excesses) is 0.038 magnitudes. 

Small color shifts in our clusters are possible due to
effects such as metallicity differences or low levels of reddening
(e.g., the probable reddening of Praesepe proposed by Taylor et
al.~(2006) translates to $\sim$ 0.01 mag at K$_S$).  We allowed the
colors for each cluster to shift horizontally (in the $K_S-[24]$
direction) to find the overall best fit to the ensemble of stars, with
different shifts allowed for x$\le$0 and x$>$0.  We found the horizontal
offset for each cluster by minimizing the calculated $\chi^2$, defined
as:

\begin{equation}
\chi^2=\sum [ (K_S-[24])_{\rm observed}-(K_S-[24])_{\rm predicted} ]^2.
\end{equation}

\begin{figure}
\figurenum{2}
\label{fig2}
\plotone{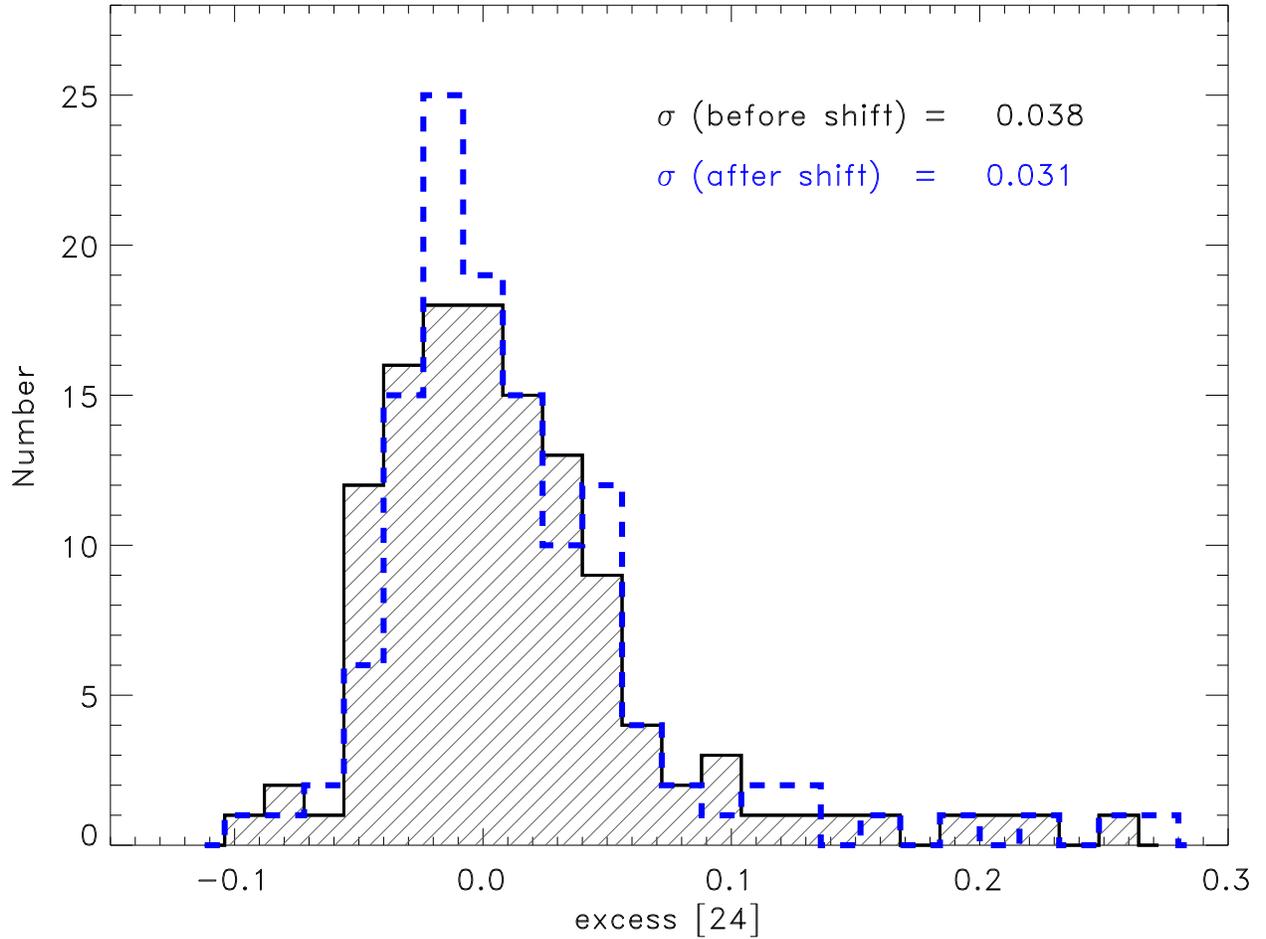}
\caption{Histogram of $V - [24]$ relative to the main sequence locus, for the stars retained in our analysis. The general nature of the distribution is similar before and after we shift the loci for the individual clusters to minimize the deviations from it. The stars with excesses are shown by the low-lying distribution extending to positive values $>$ 0.1.}
\end{figure}

\noindent
We did not want to skew the $\chi^2$ values with the redder K and M
stars; therefore, we chose an upper bound of $V-K_S=$2.5 for the
$\chi^2$ minimization. We included only a range between $-$0.1 and 0.1
in $K_S-[24]$ (i.e., 3 $\sigma$ in our fit) so that possible excess
stars did not contaminate our results. Table~\ref{shift} gives shifted
values for each of the clusters. We found a systematic offset of about 0.02
between the blue and red segments of the fit, which probably indicates a residual
problem in our determination of the main sequence locus. 
The scatter improves to 0.031 mag after implementing these shifts (Figure \ref{fig2}).

\section{Results}

\subsection{Identification of stars with excess emission}

We first identified excess candidates by requiring that the
difference between the observed $K_S-[24]$ and the predicted
$K_S-[24]$ is $\gtrsim$ 0.1 mag. There are nine stars that meet this
criterion.  To ensure the excess is significant, we
also compute $\chi_{24}$, defined as

\begin{equation} 
\chi_{24}=\frac{(K_S-[24])_{{\rm observed}}-(K_S-[24])_{{\rm predicted}}}{e_{K_S-[24]}}.
\end{equation} 

\begin{figure}
\figurenum{3}
\label{fig3}
\plotone{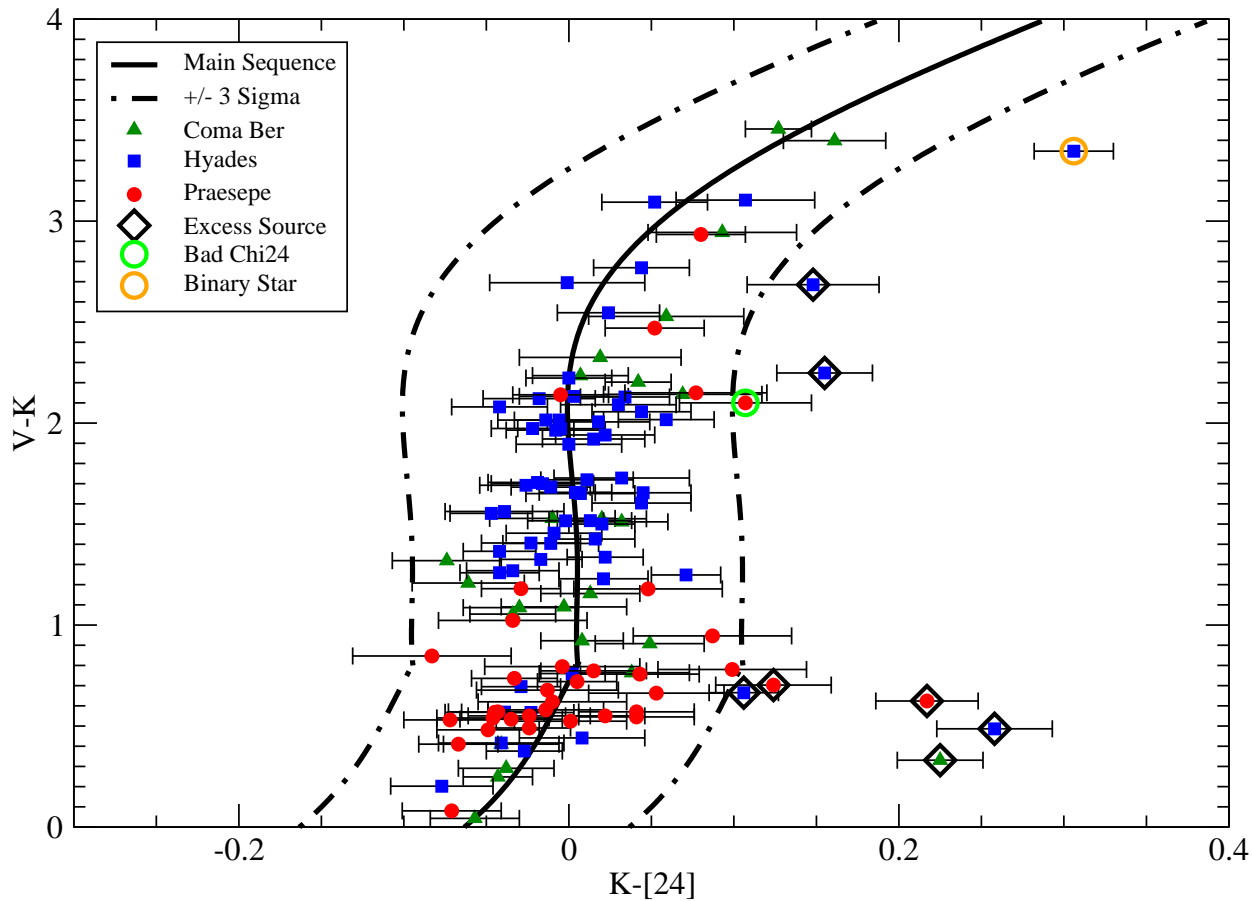}
\caption{$V-K_S$ vs. $K_S - [24]$ color-color plot for 670 Myr cluster
  members that we retain in our analysis. The solid line is the locus
  of main sequence stars and the dash-dot lines are at 3-$\sigma$
  above and below in $K_S - [24]$. The stars that pass our tests for
  having 24 $\mu$m excesses are indicated with diamonds. One
  additional star circled in green, 2MASSJ08385506+1911539, lies to the right of the
  3-$\sigma$ locus but fails our signal-to-noise criterion for a
  verified excess, as can also be seen by the large error bars for
  it (Section 3.1). HIP21179 is circled in orange because of a possible false
  excess due to a binary companion (Section 3.3).} 
\end{figure}

\noindent
A significant excess requires $\chi_{24} \geq 3$ (Su et al.\ 2006,
Trilling et al.\ 2008). One of the nine stars, 2MASS J08385506+1911539, from the Praesepe cluster 
failed this test ($\chi_{24}=$2.7); we eliminated it from the list of confirmed excesses. 
We therefore find that eight out of our sample of 122 stars have excesses at 24~$\micron$ 
greater than 10\% of their photospheric emission and at a minimum 3$\sigma$ confidence 
level (Figure \ref{fig3} and Table \ref{tab_excesses}).  Five of these
eight stars have previously been identified as having 24 $\mu$m excesses:
HD~28226 and HD~28355 (Su et al.\ 2006, Cieza et al.\ 2008);
2MASS~J08411840+1915394 (G\'asp\'ar et al.\ 2009); and HD~285690 and
HD~286789 (Stauffer et al.\ 2010).  The three stars with newly
discovered excesses are HD~108382, HIP~21179, and
2MASS~J08411840+1915394. Of these eight stars, five are from
Hyades, two are from Praesepe, and one is from Coma Ber. 

\subsection{Comparison of results from the three clusters}

To first order, we have assumed that the data for all three clusters
are homogeneous. This assumption can be tested by considering them
individually.  In addition to our overall
$\sigma=0.031$ we also calculate $\sigma$ for each individual cluster,
applying a gaussian fit to the data in each cluster separately.  We
exclude stars that are $>+3\sigma$ from the photospheric locus using the
original $\sigma$ value and obtain a new $\sigma$ value for each cluster. Given the small number of sources in the individual clusters, the differences in the scatter around the main sequence locus (see Figure 3) are not very significant. Nonetheless, we tested if the individual values would change any of our results.
For Hyades we find $\sigma$ = 0.027; however, adopting this value does not allow
us to identify any new excess sources. For Coma Ber we find $\sigma$  = 0.036; which is larger than the
original value. However, the only source from Coma Ber with excess still remains beyond this new
3 $\sigma$ threshold. Finally, for Praesepe we find $\sigma$ = 0.032, which is similar to the overall average for
the three clusters. We suspect that the differences in $\sigma$ are due to small number statistics, but in any case they do not affect our conclusions.

\subsection{Possible False Detections}

A general possibility for false detections is confusion with background infrared-emitting galaxies. For this situation, we have adapted the analysis of G\'asp\'ar et al. (2009), adjusting for our smaller matching radius (1\farcs5 vs. 3\farcs6). Because faint and distant galaxies are uniformly distributed over the sky, the analysis should apply to all three clusters. We find that there is a 10\% probability of a single chance coincidence in our entire sample, and a 5\% probability of two coincidences. We therefore make no corrections for such matches. 

Plavchan et al.\ (2009) discuss the possibility that an M dwarf
spectroscopic binary companion to a K dwarf may produce 10\%
``excess'' (superphotospheric) flux at 24 $\micron$.  This would
produce an artificial excess (false positive) source in our
color-color space.  There is only one known binary K star in our
sample that shows an excess: HIP~21179 (Bender \& Simon\ 2008).  This
star is considered to be a secure member of the Hyades (Perryman et
al.\ 1998) with a relatively high X-ray luminosity (Stern et
al.\ 1995). The masses of the components are $0.79 \pm 0.08$ M$_\odot$
and $0.58 \pm 0.15$ M$_\odot$ (Bender \& Simon 2008). It is variable (V1147 Tau) with an amplitude in the visible of $\sim$ 0.17 mag (Watson et al. 2011). The HIP~21179 excess is 6$\sigma$ and 19\% higher than the photospheric prediction, but given this list of issues we do not consider the presence of a debris disk to be secure and we discard it from our sample of disks.

The second late-type star with an apparent excess is HD 285690. It is also a secure Hyades member (Perryman et al.\ 1998), but has no evidence for binarity and is of low X-ray luminosity (Stern et al. 1995). Although it is variable (V985 Tau), the amplitude is only 0.02 mag (Watson et al.\ 2011), not at a level that significantly undermines the identification of the 24 $\mu$m excess. This star is therefore likely to be a genuine debris disk detection. The third such star, HD 286789, is also a secure Hyades member with low X-ray luminosity. It shows variability of 0.06 magnitudes in the visible (Watson et al.\ 2011), but given the general reduction in variability amplitude for late dwarf stars in the infrared (e.g., Koen et al.\ 2005), the excess is reasonably secure. Since the $\chi^2$ value for the excess of HD 286789 is only slightly above 3, we combined the 2MASS H and K measurements (the error on J is large) to compute an equivalent K of 7.792 $\pm$ 0.022, and a new $\chi^2$ of 3.7. Both of these stars are sufficiently bright at 24 $\mu$m that the probability of chance coincidences with background galaxies that might create false excesses is small: $\sim$ 1\% for one or more such coincidences in our entire sample (scaling from G\'asp\'ar et al.\ 2009). We therefore retain both stars in our sample.

\subsection{Comparison to Previous Work}

Our work is the first analysis of the {\it Spitzer} MIPS data for Coma
Ber; however, the observations of the Hyades have been analyzed
previously by Cieza et al. (2008) and Stauffer et al. (2010), while a
paper on Praesepe has been published by G\'asp\'ar et al. (2009). In
the Hyades, Cieza et al. (2008) found only a single star with a
significant excess, HD 28355; this star is also found to have an
excess by Su et al. (2006) and in our work. The identification of five
additional stars with excesses by Stauffer et al. (2010) and by us results directly from the smaller RMS
scatter in the data reduction and our more accurate extrapolation of the photospheric emission to 24 $\mu$m. Stauffer et al. (2010) found an excess for one
case not in our sample, the F8V star, HD~26784. We agree that it has a nominal excess of $\sim$ 0.09 mag (K$_S$ = 5.862, [24] = 5.782, shift of -0.014), near the threshold for detection, but it fell below our uniform threshold for an excess of 0.1 mag. The largest term in the uncertainty for this star is the 2MASS K magnitude. We have augmented it by using standard JHK colors (Tokunaga 2000, Carpenter et al. 2008, with the correction described in Rieke et al. 2008) appropriate for its V-K color (=1.26) to compute equivalent K magnitudes from 2MASS J and H, assuming an additional uncertainty of 0.01 for J-K and 0.005 for H-K. A weighted average of the three measures yields K = 5.862 $\pm$ 0.012, and confirms that an excess is detected for this star at a statistically significant level.

G\'asp\'ar et al. (2009) identified four stars with probable excesses
in Praesepe; three of them (numbers 77, 134, and 181) do not pass the
signal to noise threshold for inclusion in our sample. The remaining
star, number 143 = 2MASSJ08411840+1915394, is also found to have an excess in our study, while
our measurement of number 134 = 2MASS08410961+1951186 verifies its excess, even if its error in $K_S - [24]$ was too large to include it in our sample. The remaining two sources, numbers 77 = 2MASS08395998+1934405  and 181 = 2MASS08424021+1907590, do not pass our criteria for firmly-established excesses because our noise estimation is significantly more conservative (larger noise) than that used by G\'asp\'ar et al. (2009).  

\subsection{Secondary sample}

The case of HD 26784 suggests that there may be other plausible
excesses below our overall threshold of $K_S$ - [24] $>$ 0.1 and [24]
error $\le$ 0.05. In fact, a search relaxing the requirements on size
of excess and K$_S$ - [24] error yields Melotte 111 AV573 =
2MASS12133585+2910216 from Coma Ber as having evidence for a debris disk.   

We divide the detected excesses into two samples. The primary one is
defined by our initial criteria (excess at $>$ 3 $\sigma$ significance
and $>$ 0.1 magnitudes), and can be compared directly with previous
studies identifying excesses by similar criteria. The secondary sample
consists of HD 26784 (F8V), Melotte 111 AV573 ($\sim$ K0), and
2MASS08410961+1951186 ($\sim$ G4 from its V-K color). 
%A general
%possibility for false detections is confusion with background infrared-emitting
%galaxies. For this situation, we have adapted the analysis of
%G\'asp\'ar et al. (2009), adjusting for our smaller matching radius
%(1\farcs5 vs. 3\farcs6).  Because faint and distant galaxies are uniformly
%distributed over the sky, the analysis should apply to all three
%clusters. We find that there is a 10\% probability of a single chance
%coincidence with a galaxy bright enough to produce a false excess in
%our entire sample, and a 5\% probability of two coincidences. We
%therefore make no corrections for such matches.

%The first of these stars is sufficiently bright that there is virtually no chance of a chance coincidence with a background galaxy; at the brightness of the second, we expect a 10\% possibility for one coincidence in the entire sample, and for the third a 5\% possibility. 

\section{Discussion}

From our study, the overall rate of 24-$\mu$m excesses $>$ 10\% of the photospheric level at 670~Myr is
$5.7^{+3.1}_{-1.7}$\% (7/122 sources). The errors (here and in the following) 
are calculated using a
binomial distribution for small number statistics. Our result is higher than the excess rate for Praesepe of
2.1$^{+4.1}_{-2.1}$\% (4/193~sources) found by
G\'asp\'ar et al.\ (2009), but the difference can be largely explained by the smaller threshold for identifying an excess in our study (0.1 magnitudes instead of 0.15); in any case, the two values agree within the errors.

We now compare the evolution as a function of stellar type. A
first-order estimate of the time scale for debris disk evolution is
given by Wyatt (2008), equation (16). To apply this result, we assume
an average luminosity ratio of 8 between our early-type (B9 - F4) and
solar-like (F5 - K9) samples, and a mass ratio of 1.7. The thermal
equilibrium distance from the fiducial stars is then different by a
factor of 2.8. We assume that typical planetesimal disk masses scale
with the mass of the star (Natta et al.\ 2000). We find that the time scale is dominated by the strong radial dependence in equation (16), which indicates that the decay around the early-type stars should be an order of magnitude slower than around the solar-like ones.

Our sample is ideally suited to determine the excess rate for stars near the mass of the sun and
with well-constrained ages near 670 Myr. We used spectral types from SIMBAD and placed stars
without a given spectral type from SIMBAD in bins according to their $V-K_S$ colors using Table
3 in Koornneef (1983). We find that the excess rate for F5--K9 type stars is $2.7^{+3.3}_{-1.7}$\% (2/75 stars), in agreement with the value of 1.9 $\pm$ 1.2 \% from G\'asp\'ar et al. (2009). However, the secondary sample includes three well-detected debris disks around stars in the same spectral range, so this value may be a modest underestimate. We can compare it with that obtained by Sierchio et al.~(2010), who
measured the excess rate for stars in the same spectral range in the 115 Myr Pleiades cluster and the similar-age Blanco 1
association using the same reduction method and excess threshold of 10\%. They report an excess
rate of $31.5^{+5.2}_{-4.5}$\% at 24 \micron\ (28/89). We can add to
their sample observations in the $\alpha$ Per cluster (Carpenter et
al.\ 2009) at an age of 85 Myr, for a total of $29.4^{+4.9}_{-4.1}$\%
(30/102) at an age of $\sim$ 100 Myr.

We have assembled similar data for B9--F4 stars from the literature.  For the age range 50--99 Myr there are
17/47 stars with 24 \micron\ excesses, or $36.2^{+7.3}_{-6.2}$\% (Rieke et al. 2005; Su et al.\ 2006; Siegler et al.\ 2007; Balog et al.\ 2009), while in the
100--199 Myr range there are 33/93, or $35.5^{+5.1}_{-4.6}$\% (Rieke et al.\ 2005; Su et al.\ 2006; Gorlova et al.\ 2006). The combined result for 50 - 199 Myr is $35.7^{+4.3}_{-3.8}$\% (50/140). In the 400--1000
Myr range, the corresponding number is 6/77, or $7.8^{+4.2}_{-2.1}$\%
(Su et al. 2006; this work). This value agrees well with that in
G\'asp\'ar et al. (2009) of 6.5 $\pm$ 4.1 \%.

The incidence of debris disks at $\sim$ 100 Myr shows no difference
between the two ranges of spectral type. At $\sim$ 670 Myr, the later
spectral types show a lower incidence, but the two values still agree
at the 1.5 $\sigma$ level. Moreover, the detection of three later-type stars in the secondary sample (of which HD 26784 only barely misses the primary sample) suggests that a significant number of such stars retain debris disks at this age. This conclusion has been reached previously (e.g., Trilling et al.\ (2008); G\'asp\'ar et al.\ (2009); Carpenter et al.\ (2009)). 

However, all previous such works have depended in part or entirely on
samples of field stars. Given the uncertainties in age estimates for
these stars, combined with the rapid decline in the incidence of
excesses (from $\sim$ 30\% at 100 Myr to $\sim$ 3 \% at 670 Myr), age
errors that place a small number of young stars among the older part
of a sample could yield the observed number of excesses. Our work
removes this source of uncertainty by basing the result entirely on
observations of cluster members with well-determined ages. Thus, the
data appear to contradict the first-order theoretical time scale
difference for disk decay as a function of stellar type.

\section{Conclusion}

We used 24~$\micron$ {\it Spitzer} observations of stars in three clusters
(Hyades, Coma Ber, and Praesepe)
to measure the incidence of debris disks at $\sim$
670~Myr and over a broad range of spectral types. The combination of three clusters at the same
age gives a better representation of the excess behavior because
we can use the closest cluster (Hyades) to detect excesses around low mass
stars, while having access to significant numbers of higher mass
stars in the more distant cluster (Praesepe). 
We compared the V-K$_S$, K$_S$-[24] of the cluster members
with the color locus of 1300 field stars. 
The dispersion around this locus is 10\% (3$\sigma$), which
we adopt as the threshold for identification of an excess.

We find an overall excess rate of $5.7^{+3.1}_{-1.7}$\% for stars at
$\sim$ 670 Myr. This value shows substantial decay from the rates of
$29.4^{+4.9}_{-4.1}$\% for F5 - K9 stars and $35.7^{+4.3}_{-3.8}$\%
for B9 - F4 stars at $\sim$ 100 Myr. However, the decay appears to be
similar within the errors between these ranges of spectral types. This result is contrary to first-order estimates, which would indicate an order-of-magnitude slower decay for the early type stars than the solar-like ones. 

\acknowledgements
This work was supported at Northern Arizona University by funding from
the Spitzer Science Center/JPL. Funding was also provided by the
National Science Foundation through a Research Experience for
Undergraduates (REU) position at Northern Arizona University.  Support
was also provided by contract 1255094 from Caltech/JPL to the
University of Arizona. This research made use of the SIMBAD and Vizier
database, operated at CDS, Strasbourg, France. This work also uses
data products from the Two Micron All Sky Survey, which is a joint
project of the University of Massachusetts and the Infrared Processing
and Analysis Center/California Institute of Technology, funded by the
National Aeronautics and Space Administration and the National Science
Foundation. We would also like to thank Jennifer Sierchio and Andras
G\'asp\'ar for helpful comments and discussions and the anonymous
referee whose suggestions improved this paper.

\clearpage
\bibliographystyle{apj}
\bibliography{ref}

\begin{thebibliography}
\expandafter\ifx\csname natexlab\endcsname\relax\def\natexlab#1{#1}\fi

\bibitem[Abad \& Vicente(1999)]{1999yCat..41360307A} Abad, C., \& Vicente, B.\ 1999, VizieR Online Data Catalog, 413, 60307 

\bibitem[Agnor et al.(1999)]{1999Icar..142..219A} Agnor, C.~B., Canup, 
R.~M., \& Levison, H.~F.\ 1999, \icarus, 142, 219

\bibitem[Aumann et al.(1984)]{1984ApJ...278L..23A} Aumann, H.~H., et al.\ 
1984, \apjl, 278, L23 


\bibitem[Balog et al.(2009)]{}Balog, Z., et al. 2009, \apj,698, 1989 

\bibitem[Bender \& Simon(2008)]{2008ApJ...689..416B} Bender, C.~F., \& Simon, M.\ 2008, \apj, 689, 416 


\bibitem[Carpenter et al.(2008)]{} Carpenter, J. M. et al. 2008, \apjs, 
179, 423

\bibitem[Carpenter et al.(2009)]{}Carpenter, J. M., et al. 2009, \apjs, 181, 197

\bibitem[Chambers(2001)]{2001Icar..152..205C} Chambers, J.~E.\ 2001, 
\icarus, 152, 205

\bibitem[Cieza et al.(2008)]{} Cieza, L. A., Cochran, W. D., \& 
Augereau, J.-C. 2008, \apj, 679, 720

\bibitem[Collier et al.(2009)]{} Collier, C. A. et al. (2009), \mnras, 400, 451

\bibitem[Cutri et al.(2003)]{} Cutri, R. M. et al. (2003), The IRSA 2MASS All-Sky Point Source Catalog, NASA/IPAC

\bibitem[Decin et 
al.(2000)]{2000A&A...357..533D} Decin, G., Dominik, C., Malfait, K., Mayor, M., \& Waelkens, C.\ 2000, \aap, 357, 533 

\bibitem[Decin et al.(2003)]{2003ApJ...598..636D} Decin, G., Dominik, C., 
Waters, L.~B.~F.~M., \& Waelkens, C.\ 2003, \apj, 598, 636 

\bibitem[De Gennaro et al.(2009)]{} De Gennaro, S., von Hippel, T., Jeffreys, W. H., Stein, N., van Dyk, D., 
\& Jeffrey, E. 2009, \apj, 696, 12

\bibitem[Dominik 
\& Decin(2003)]{2003ApJ...598..626D} Dominik, C., \& Decin, G.\ 2003, \apj, 598, 626 

\bibitem[Engelbracht et al.(2007)]{} Engelbracht, C. M. et al. 2007,
  \pasp, 119, 994

\bibitem[Hahn 
\& Malhotra(1999)]{1999AJ....117.3041H} Hahn, J.~M., \& Malhotra, R.\
1999, \aj, 117, 3041

\bibitem[G{\'a}sp{\'a}r et al.(2009)]{2009ApJ...697.1578G} G{\'a}sp{\'a}r, 
A., Rieke, G.~H., Su, K.~Y.~L., Balog, Z., Trilling, D., Muzzerole, J., 
Apai, D., \& Kelly, B.~C.\ 2009, \apj, 697, 1578 

\bibitem[Gomes et al.(2004)]{2004Icar..170..492G} Gomes, R.~S., Morbidelli, 
A., \& Levison, H.~F.\ 2004, \icarus, 170, 492

\bibitem[Gorlova et al.(2006)]{2006ApJ...649.1028G} Gorlova, N., Rieke, 
G.~H., Muzerolle, J., Stauffer, J.~R., Siegler, N., Young, E.~T., 
\& Stansberry, J.~H.\ 2006, \apj, 649, 1028 

\bibitem[Gordon et al.(2005)]{} Gordon, K. D. et al. 2005, \pasp, 117, 503

\bibitem[Habing et 
al.(2001)]{2001A&A...365..545H} Habing, H.~J., et al.\ 2001, \aap, 365, 545 

\bibitem[Haisch et al.(2001)]{2001ApJ...553L.153H} Haisch, K.~E., Jr., 
Lada, E.~A., \& Lada, C.~J.\ 2001, \apjl, 553, L153 

\bibitem[Kenyon \& Bromley(2004)]{} Kenyon, S. J., \& Bromley, B. C. 2004, \apjl, 602, 133

\bibitem[Koen et al.(2005)]{} Koen, C., Tanab\'e, T., Tamura, M., \& Kusakabe, N. 2005, \mnras, 362, 727

\bibitem[Koerner et al.(2010)]{2010ApJ...710L..26K} Koerner, D.~W., et al.\ 2010, \apjl, 710, L26 

\bibitem[Koornneef(1983)]{1983A&AS...51..489K} Koornneef, J.\ 1983, \aaps, 51, 489 

\bibitem[Lagrange et al.(2000)]{2000prpl.conf..639L} Lagrange, A.-M., 
Backman, D.~E., \& Artymowicz, P.\ 2000, Protostars and Planets IV,
639 

\bibitem[Levison et al.(2007)]{2007prpl.conf..669L} Levison, H.~F., 
Morbidelli, A., Gomes, R., 
\& Backman, D.\ 2007, Protostars and Planets V, 669

\bibitem[Meyer et al.(2008)]{2008ApJ...673L.181M} Meyer, M.~R., et al.\ 
2008, \apjl, 673, L181 

\bibitem[Morales et al.(2011)]{2011ApJ...730L..29M} Morales, F.~Y., Rieke, G.~H., Werner, M.~W., Bryden, G., Stapelfeldt, K.~R., 
\& Su, K.~Y.~L.\ 2011, \apjl, 730, L29 

\bibitem[Morishima et al.(2010)]{2010Icar..207..517M} Morishima, R., 
Stadel, J., \& Moore, B.\ 2010, \icarus, 207, 517

\bibitem[Natta \& Mannings(2000)]{}Natta, A., Grinin, V., \& Mannings, V. 2000, Protostars \& Planets IV (Tucson: U. Ariz. 
	Press), p. 559


\bibitem[Paulson et al.(2004)]{2004AJ....127.3579P} Paulson, D.~B., 
Cochran, W.~D., \& Hatzes, A.~P.\ 2004, \aj, 127, 3579 

\bibitem[Perryman et 
al.(1997)]{1997A&A...323L..49P} Perryman, M.~A.~C., et al.\ 1997, \aap, 323, L49 

\bibitem[Perryman et 
al.(1998)]{1998A&A...331...81P} Perryman, M.~A.~C., et al.\ 1998, \aap, 331, 81 

\bibitem[Plavchan et al.(2009)]{2009ApJ...698.1068P} Plavchan, P.,
  Werner, M.~W., Chen, C.~H., Stapelfeldt, K.~R., Su, K.~Y.~L.,
  Stauffer, J.~R., \& Song, I.\ 2009, \apj, 698, 1068 

\bibitem[Raymond et al.(2006)]{2006Icar..183..265R} Raymond, S.~N., Quinn, 
T., \& Lunine, J.~I.\ 2006, \icarus, 183, 265

\bibitem[Rhee et al.(2007)]{2007ApJ...660.1556R} Rhee, J.~H., Song, I., 
Zuckerman, B., \& McElwain, M.\ 2007, \apj, 660, 1556 

\bibitem[Rieke et al.(2004)]{2004ApJS..154...25R} Rieke, G.~H., et al.\ 
2004, \apjs, 154, 25 

\bibitem[Rieke et al.(2005)]{2005ApJ...620.1010R} Rieke, G.~H., et al.\ 
2005, \apj, 620, 1010 

\bibitem[Rieke et al.(2008)]{2008AJ....135.2245R} Rieke, G.~H., et al.\ 
2008, \aj, 135, 2245 

\bibitem[Siegler et al.(2007)]{}Siegler, N. et al. 2007, \apj, 654, 580

\bibitem[Sierchio et al.(2010)]{2010ApJ...712.1421S} Sierchio, J.~M., 
Rieke, G.~H., Su, K.~Y.~L., Plavchan, P., Stauffer, J.~R., 
\& Gorlova, N.~I.\ 2010, \apj, 712, 1421 

\bibitem[Spangler et al.(2001)]{2001ApJ...555..932S} Spangler, C., Sargent, A.~I., Silverstone, M.~D., Becklin, E.~E., 
\& Zuckerman, B.\ 2001, \apj, 555, 932 

\bibitem[Stauffer et al.(2010)]{2010ApJ...719.1859S} Stauffer, J.~R., et 
al.\ 2010, \apj, 719, 1859 

\bibitem[Stern et al.(1995)]{1995ApJ...448.683S} Stern, R. A., Schmitt, 
J. H. M. M., \& Kahabka, P. T.\ 1995, ApJ, 448, 683

 

\bibitem[Su et al.(2006)]{2006ApJ...653..675S} Su, K.~Y.~L., et al.\ 2006, \apj, 653, 675 

\bibitem[Su et al.(2010)]{} Su, K. Y. L. et al. (2010)]{} Su, K. Y. L. et al. 2010, AAS Meeting 215, abs. 428.26

\bibitem[Taylor(2006)]{2006AJ....132.2453T} Taylor, B.~J.\ 2006, \aj, 132, 2453 

\bibitem[Tokunaga(2000)]{} Tokunaga, A. T. 2000, in Allens 
Astrophysical Quantities, 4th Ed. (New York: AIP Press), p 143.

\bibitem[Trilling et al.(2008)]{2008ApJ...674.1086T} Trilling, D.~E., et 
al.\ 2008, \apj, 674, 1086 

\bibitem[Watson et al.(2011)]{} Watson, C., Henden, A. A., \& Price, A. 2011, AAVSO Variable Star Index (online at VizieR)

\bibitem[Wyatt(2008)]{2008ARA&A..46..339W} Wyatt, M.~C.\ 2008, \araa, 46, 339 

%%%%%%%%%%%%%%%%%%%%%















\end{thebibliography}

\clearpage

\begin{deluxetable}{llllllllllll}
\tablewidth{0pt}
\tabletypesize{\scriptsize}
\tablecaption{Cluster Stars Considered\label{Cluster Stars Considered}}
\tablehead{\colhead{Name} & \colhead{Sp} &\colhead{V}&\colhead{K$_S$}&\colhead{K$_{err}$} & \colhead{f24} & \colhead{[24]}&\colhead{[24]$_{err}$} & \colhead{(K-[24])$_{err}$} & \colhead{Incl?} & \colhead{Excl.} & \colhead{Cluster}\\ \colhead{} & \colhead{Type} &\colhead{}&\colhead{}&\colhead{} & \colhead{} & \colhead{}&\colhead{} & \colhead{} & \colhead{}& \colhead{note} & \colhead{}}
\startdata
   Cl* Melotte 111 AV1119 &        G5V &   10.830 &    9.295 &    0.019 &    1.453 &    9.233 &    0.057 &    0.060 &   n &        4 &   c \\
   Cl* Melotte 111 AV1183 &        K3V &   11.500 &    8.972 &    0.020 &    1.932 &    8.924 &    0.043 &    0.047 &   y &   \nodata&   c \\
   Cl* Melotte 111 AV1280 &       K4V: &   11.720 &    9.476 &    0.018 &    1.164 &    9.474 &    0.068 &    0.070 &   n &    1,2,4 &   c \\
   Cl* Melotte 111 AV1284 &         F2 &    4.780 &    3.236 &    0.244 &  395.500 &    3.146 &    0.011 &    0.244 &   n &        4 &   c \\
   Cl* Melotte 111 AV1288 &         G0 &   12.520 &   10.519 &    0.019 &   \nodata&   \nodata&    0.011 &    0.022 &   n &        3 &   c \\
   Cl* Melotte 111 AV1292 &        F8V &   10.990 &    9.606 &    0.021 &    1.047 &    9.589 &    0.077 &    0.080 &   n &        4 &   c \\
   Cl* Melotte 111 AV1293 &       G5V: &   12.580 &   10.796 &    0.024 &    0.359 &   10.750 &    0.132 &    0.135 &   n &        4 &   c \\
   Cl* Melotte 111 AV1297 &        G8V &   10.440 &    8.752 &    0.021 &    2.418 &    8.680 &    0.070 &    0.073 &   n &        4 &   c \\
\enddata
\tablecomments{The characteristics of all stars in our sample.  A
  total of 355 are presented here.  Notes for column for reason for
  exclusion are as follows: (1) Contamination from nearby stars (2)
  FWHM much greater or much less than 5.5 arcsec (3) Signal to
  Noise$<$3 (4) K-[24] error greater than .05 (5) No V mag (6) No 24
  $\mu$m counter part within 1\farcs5 search radius (7) Giant stars.} 
\end{deluxetable}																														
\begin{deluxetable}{ccc}
\tablewidth{0pt}
\tablecaption{Derived $K_S-[24]$ shifts in each of the clusters\label{shift}}
\tablehead{ 	
\colhead{Cluster} & \colhead{Stars with $V-K_S \leq 0.8$} & \colhead{Stars with $V-K_S >0.8$} 
}
\startdata
Hyades   &  $+$0.011 & $-$0.009   \\
Coma Ber &  $+$0.028 & $+$0.011  \\
Praesepe &  $+$0.009 & $-$0.014  \\
\enddata 
\end{deluxetable}

\begin{deluxetable}{llcccll}
\tablewidth{0pt}
\tabletypesize{\small}
\tablecaption{Identified 24 $\mu$m excess sources\label{tab_excesses}}
\tablehead{ 
\colhead{Name} & \colhead{Type} & \colhead{excess\tablenotemark{a}} & \colhead{error\tablenotemark{b}}
& \colhead{$\chi_{24}$} &\colhead{Cluster} & \colhead{Previous ID} 
}
\startdata
                 HD028226 &  Am  &  0.109 &  0.021  &   5.2 & Hyades & Su et al. (2006)     \\
                 HD028355 &  A7V &  0.273 &  0.035  &   7.9 & Hyades & Su et al. (2006)    \\
                 HD285690 &  K0  &  0.155 &  0.029  &   5.3 & Hyades & Stauffer et al. (2010)    \\
                 HD286789 &  K7  &  0.127 &  0.040  &   3.7 & Hyades & Stauffer et al. (2010)    \\
                 HIP21179 &  K0  &  0.188 &  0.024  &   7.7 & Hyades & none\tablenotemark{c}    \\
                 HD108382 &  A4V &  0.252 &  0.026  &   9.5 & Coma Ber & none    \\
%   2MASSJ08384695+1930033 &  0.095 &  0.045  &   2.1 & Praesepe & none     \\
   2MASSJ08385506+1911539 &  K3V & 0.108 &  0.040  &   2.7 & Praesepe & none\tablenotemark{d}    \\
   2MASSJ08403296+1911395 &  F0V &  0.125 &  0.035  &   3.6 & Praesepe & none    \\
\enddata
\tablenotetext{a}{$(K_S-[24])_{{\rm observed}}-(K_S-[24])_{{\rm
      predicted}}$, in magnitudes}
\tablenotetext{b}{Error calculated using Equation 1}
\tablenotetext{c}{Excess probably not associated with a debris disk;
  see Section 3.3.}
\tablenotetext{d}{Fails $\chi_{24}$ test, see Section 3.1.}
\tablenotetext{e}{Secondary sample; see Sections 3.4 \& 3.5.} 

\end{deluxetable}

%\begin{figure}
%\figurenum{4}
%\label{fig4}
%\plotone{DecayPlot.eps}
%\caption{The decay of the excess rate from $\sim$100 Myr to $\sim$670 Myr for the
%  spectral ranges B9-F4 and F5-K9.  The blue squares represent the
%  average excess rate combining samples discussed in Section 4 for
 % the early spectral types, B9-F4. The average excess rate drops from
 % $35^{+4.3}_{-3.8}$\% to $7.8^{+4.2}_{-2.1}$\% in this spectral range. The red circles show the average
 % excess rate combining different samples also discussed in Section 4
 % for later spectral types, F5-K9.  The average excess rate drops from
 % $29.4^{+4.9}_{4.1}$\% to $2.7^{+3.3}_{-1.7}$\% for this spectral range. 
 % The lines connecting
 % the points have similar slopes representing that the decline of disk incidence is not a
 % function of spectral type for warm dust probed by 24 $\micron$ data.}
%\end{figure}

\end{document}